\documentclass[aps,prl, superscriptaddress, twocolumn]{revtex4-2}
\usepackage{natbib}
\usepackage{graphicx}
\usepackage{bm}
\usepackage[utf8]{inputenc}
\usepackage{color,soul}
\usepackage{hyperref}
\usepackage{float}
\usepackage{csquotes}
\usepackage{color,soul}
\hypersetup{colorlinks=true, urlcolor=blue, citecolor=blue}
\usepackage{braket}
\usepackage{color}
\usepackage{amsmath}
\usepackage{comment}
\usepackage{rotating}
\usepackage{mhchem}
\usepackage{amsmath}

\begin{document}
\title{Chirality-Induced Spin Selectivity: Nonlinear
Spin Response from Electron–Phonon Scattering}
\author{Mayank Gupta \footnotemark[1]}
\affiliation{Department of Materials Science and Engineering, University of Wisconsin-Madison, WI, 53706, USA}
\author{Andrew Grieder \footnotemark[1]}
\thanks{M.G. and A.G. contributed equally.}
\affiliation{Department of Materials Science and Engineering, University of Wisconsin-Madison, WI, 53706, USA}
\author{Mayada Fadel}
\affiliation{Department of Materials Science and Engineering, Rensselaer Polytechnic Institute, 110 8th Street, Troy, New York 12180, United States} 
\author{Jacopo Simoni}
\affiliation{Department of Materials Science and Engineering, University of Wisconsin-Madison, WI, 53706, USA}
\author{Junting Yu}
\affiliation{Department of Materials Science and Engineering, University of Wisconsin-Madison, WI, 53706, USA}
\author{Ravishankar Sundararaman}
\email{sundar@rpi.edu}
\affiliation{Department of Materials Science and Engineering, Rensselaer Polytechnic Institute, 110 8th Street, Troy, New York 12180, United States} 
\author{Yuan Ping}
\email{yping3@wisc.edu}
\affiliation{Department of Materials Science and Engineering, University of Wisconsin-Madison, WI, 53706, USA}
\affiliation{Department of Physics, University of Wisconsin-Madison, WI, 53706, USA}
\affiliation{Department of Chemistry, University of Wisconsin-Madison, WI, 53706, USA}
\date{\today}
\begin{abstract}
Chirality-induced spin selectivity (CISS) generates spin-polarized 
currents in nonmagnetic materials from structural chirality alone, yet 
its microscopic origin remains debated. Using a first-principles 
spatiotemporal density-matrix dynamics approach including electron-phonon scatterings with self-consistent spin-orbit 
coupling (SOC), we elucidate the interplay of SOC, structural chirality, and spin-dependent electron-phonon interactions in driving the generation and transport of spin and orbital angular momentum.
In particular we quantitatively distinguish CISS from the collinear Edelstein effect 
(CEE) in trigonal selenium, a prototypical chiral solid. CEE yields a spatially uniform spin polarization scaling 
linearly with applied field ($S_z \propto E$). In contrast, explicit 
spin-dependent electron-phonon scattering produces a nonlinear response 
($S_z \propto E^2$) and a length-dependent spin accumulation---the 
hallmark experimental signature of CISS. We identify intervalley 
scattering mediated by chiral phonon angular momentum as the microscopic 
origin of this nonlinearity. 

\end{abstract}

\maketitle
Chirality governs a rich class of 
quantum transport and optical phenomena in condensed matter 
physics~\cite{annurevYan,YU20231}. Among these, the 
chirality-induced spin selectivity (CISS) effect converts unpolarized  charge currents into spin-polarized ones through structural chirality alone, without magnetic fields or ferromagnetic contacts~\cite{CISS_rev}. CISS has been observed across a broad class of materials---from DNA~\cite{science_dna,PRL_goldDNA} and chiral 
organic--inorganic perovskites~\cite{mat_horizon_perovs,wang_chirality,
Valy2019} to chiral crystals and 
molecules~\cite{Furukawa2017,Qian2022,naaman2019chiral}---establishing it as a general mechanism for spin generation in nonmagnetic systems~\cite{annurev_sensor}.


Despite growing experimental evidence~\cite{CISSWittmann,CISSAdhikari,
Qian2022,CISSHannah,CISSSafari}, the microscopic origin of CISS remains 
debated. SOC, structural chirality, and chiral phonons have been proposed 
as key ingredients enabling angular momentum transfer to 
electrons~\cite{Fransson_PRB2020,Fransson_PRR2023}, yet a quantitative 
first-principles treatment of spin-dependent quantum transport is still 
absent~\cite{Chiralphonon1,Chiralphonon2,CPASS}. Separately, 
current-driven spin-orbit effects in non-centrosymmetric systems---such 
as the collinear Edelstein effect (CEE)~\cite{PRL_spintex}, which 
generates spin polarization via spin--momentum locking under an applied 
field~\cite{Edelstein1990,REE_prl2015,Calavalle2022}---can produce 
signatures phenomenologically similar to CISS. Whether and how CISS and CEE 
can be distinguished in chiral solids remains an open question both 
experimentally and theoretically.


To address this challenge, we combine first-principles density-matrix 
dynamics (FPDMD)~\cite{Xu2020,Xu2021,Habib2022,Xu2024-lw,Xu2024-ze,
simoni2025} with a Wigner-function formalism~\cite{mayada_arxiv} that 
treats coherent and incoherent transport on equal footing. This enables 
spatially and temporally resolved spin and orbital dynamics with 
\textit{ab initio} electron-phonon scatterings at realistic device 
length scales.


As a prototypical chiral solid, trigonal selenium (t-Se) offers a 
clean testbed: its simple helical structure, moderate SOC, and 
symmetry-protected spin texture along the chiral axis allow the 
contributions of SOC, structural chirality, and phonons to be 
disentangled~\cite{Se_str,Se_Saeva}. We compare coherent and 
incoherent transport regimes and quantify CISS against CEE by 
contrasting spin-independent relaxation-time scattering with explicit 
spin-dependent \textit{ab initio} electron-phonon (e-ph) scatterings.


\begin{figure}
\includegraphics[width=1.0\columnwidth]{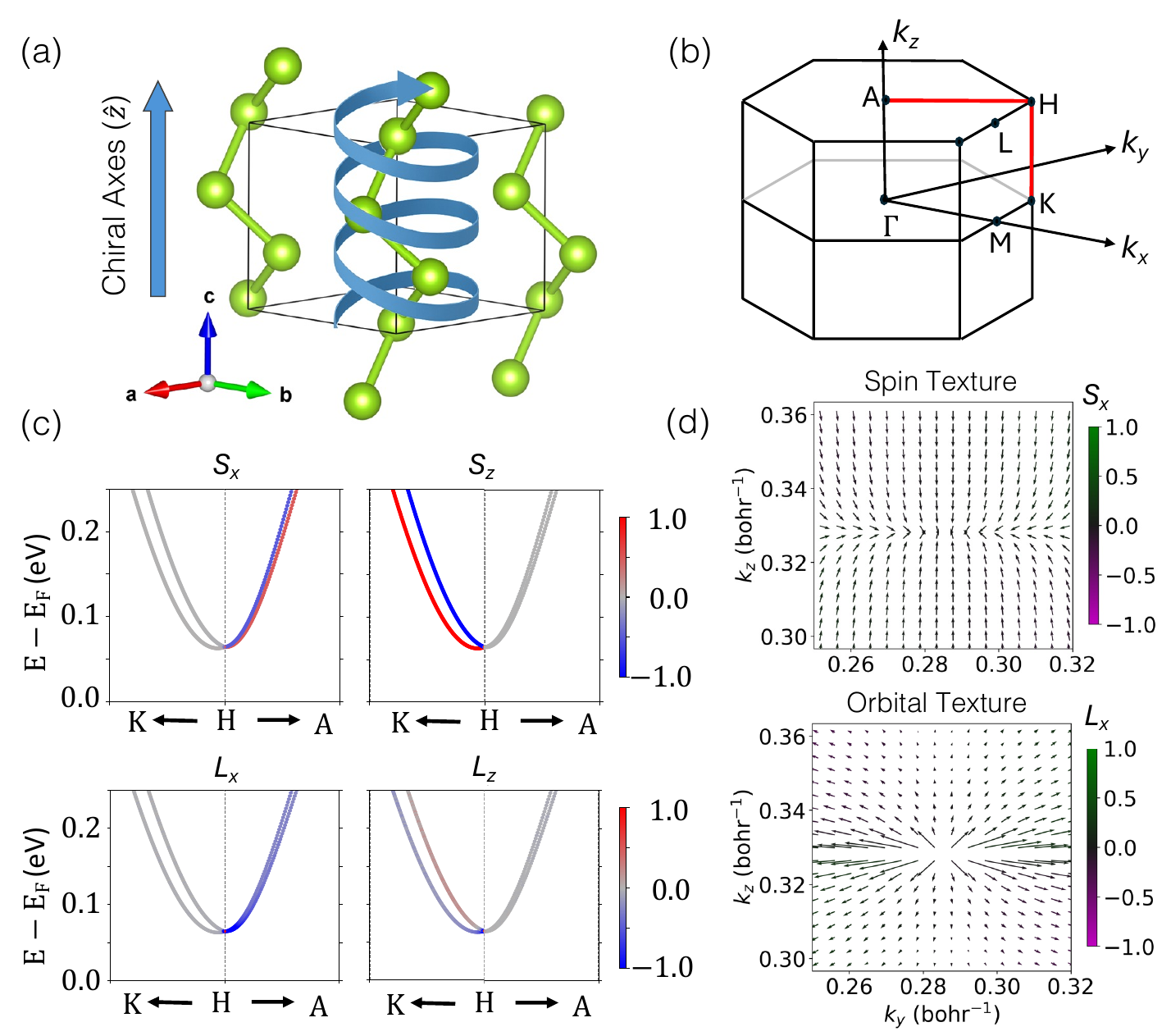}
\caption{ Structural and electronic properties of $t$-Se with left-handed chirality. (a) Crystal structure of $t$-Se, illustrating the chirality along the $\hat{z}$-axis. (b) First Brillouin zone of the crystal structure and representative high-symmetry points. The conduction band minimum is located in the vicinity of the H. (c) Spin and orbital textures are projected onto the lower conduction bands along the high-symmetry $k$-path K–H–A, as shown with red lines in the BZ. The K–H segment lies along the chiral axis, while H–A is perpendicular to it. (d) 2D spin and orbital texture of the lower conduction band centered around H in the $k_y-k_z$ plane.
}
\label{fig:str}
\end{figure}
Trigonal selenium ($t$-Se) crystallizes as helical chains along $\hat{z}$ (Fig.~\ref{fig:str}(a)), forming chiral enantiomers with space groups P3$_1$21(152) and P3$_2$21(154) for right- and left-handed forms. It belongs to point group $D_3$, featuring a threefold screw symmetry $C_3$ along the chiral axis breaking mirror and inversion symmetries and a perpendicular twofold $C_2$ axis. This symmetry breaking leads to nontrivial spin textures, which are typically radial from the H point \cite{PRL_spintex, PRL_spintex_Se}. Here, we focus on anisotropic spin and orbital angular momentum (OAM) textures (Figs.~\ref{fig:str}(c),(d)). The spin projected band structure shows that along K–H (Fig.~\ref{fig:str}(c)), only the out-of-plane spin component $S_z$ is finite due to $C_3$ symmetry, while along H–A only in-plane components $S_x$ and $S_y$ appear. This reflects symmetry-allowed SOC-induced spin splitting, yielding collinear spin–momentum locking and a quasi-persistent spin texture aligned with $\hat{z}$, with implications for spin transport discussed below (see SI for OAM details). 
%

Spin density, orbital angular momentum, and charge densities are calculated from the spatiotemporal density matrix ($\rho(\textbf{r},t)$)~\cite{Xu2021,Xu2024-ze, simoni2025}.
The equation of motion for one-particle density matrix is~\cite{mayada_arxiv}:
\begin{multline} \label{Eq:lindblad}
\dot{\rho}_{nm}^{\textbf{k}}(\textbf{r},t) = - \frac{{\bf v}_{n}({\bf k})+{\bf v}_{m}(\textbf{k})}{2}\cdot\nabla \rho_{nm}^\textbf{k} (\textbf{r},t) -\frac{i}{\hbar} [H^{\prime}, \rho]_{nm}^\textbf{k} \\ 
+\mathcal{L}\left[\rho_{nm}^\textbf{k} (\textbf{r},t)\right].
\end{multline}
Here, the first term on the right is the diffusion term, the second term is the coherent contribution driven by the external perturbation $H^\prime$. $\textbf{v}_{n}(\textbf{k})$ is the velocity of the state $\ket{n,{\bf k}}$.
The third term is the Lindbladian electron-phonon (e-ph) scattering operator given by 
\begin{multline}
\mathcal{L} \Bigl[ \rho^\textbf{k}_{n_1n_2}(\textbf{r}, t) \Bigl]
=  \frac{2\pi}{\hbar N_{\bf q}}
\sum_{{\bf q}\lambda\pm n'n'_{1}n'_{2}\textbf{k}'}n_{{\bf q}\lambda}^{\pm} \\
\times \mathrm{Re} \left[\begin{array}{c}
\left(I-\rho^\textbf{k}\right)_{n_{1}n'} A^{{\bf q}\lambda\pm}_{\textbf{k}n';\textbf{k}'n'_{1}} \rho^{\textbf{k}'}_{n'_{1}n'_{2}} A^{{\bf q}\lambda\mp}_{\textbf{k}'n'_{2};\textbf{k}n_{2}}\\
-A^{{\bf q}\lambda\mp}_{\textbf{k}n_{1};\textbf{k}'n'} \left(I-\rho^{\textbf{k}'}\right)_{n'n'_{1}} A^{{\bf q}\lambda\pm}_{\textbf{k}'n'_{1};\textbf{k}n'_{2}} \rho^{\textbf{k}}_{n'_{2}n_{2}}
\end{array}\right],
\label{eq:Lindblad}
\end{multline}

where $\pm$ indicates absorption and emission of phonons with wave vector ${\bf q} = \mp ({\bf k} - {\bf k'})$ and mode index $\lambda$, $n_{{\bf q}\lambda}^\pm \equiv n_{{\bf q}\lambda} + \frac{1}{2} \pm \frac{1}{2}$, and $N_{\bf q}$ is the total number of phonon wave vectors sampled in the Brillouin zone. $A^{{\bf q}\lambda \pm}_{{\bf k}n;{\bf k'}n'} = g_{{\bf k}n;{\bf k'}n'}^{{\bf q}\lambda\pm} \delta_G ^{1/2}(\varepsilon_n(\mathbf{k}) - \varepsilon_{n'}(\mathbf{k'}) \pm \hbar\omega_{{\bf q}\lambda}) \exp(i(\varepsilon_n(\mathbf{k}) - \varepsilon_{n'}(\mathbf{k'}))t/\hbar)$, where $g^{{\bf q}\lambda\pm}_{{\bf k}n;{\bf k'}n'}$ is the electron-phonon matrix element with SOC computed from first-principles~\cite{Xu2024-ze}, and the $\delta_G$-function is broadened to a Gaussian responsible for energy conservation, $\varepsilon_n(\mathbf{k})$ are the electronic energies~\cite{Xu2020, simoni2025}.
%
We first examine spin and orbital polarization in the ballistic regime, 
isolating the roles of SOC and structural chirality without scattering. 
We then introduce incoherent scattering in two steps: first via a 
spin-independent relaxation-time approximation (RTA) to quantify the 
CEE, then via explicit spin-dependent \textit{ab initio} e-ph 
scatterings to capture CISS. Computational details and transport setup 
are provided in Sections~I and III of the SI~\cite{SI}.

%
%

\begin{figure}
\includegraphics[width=1\columnwidth]{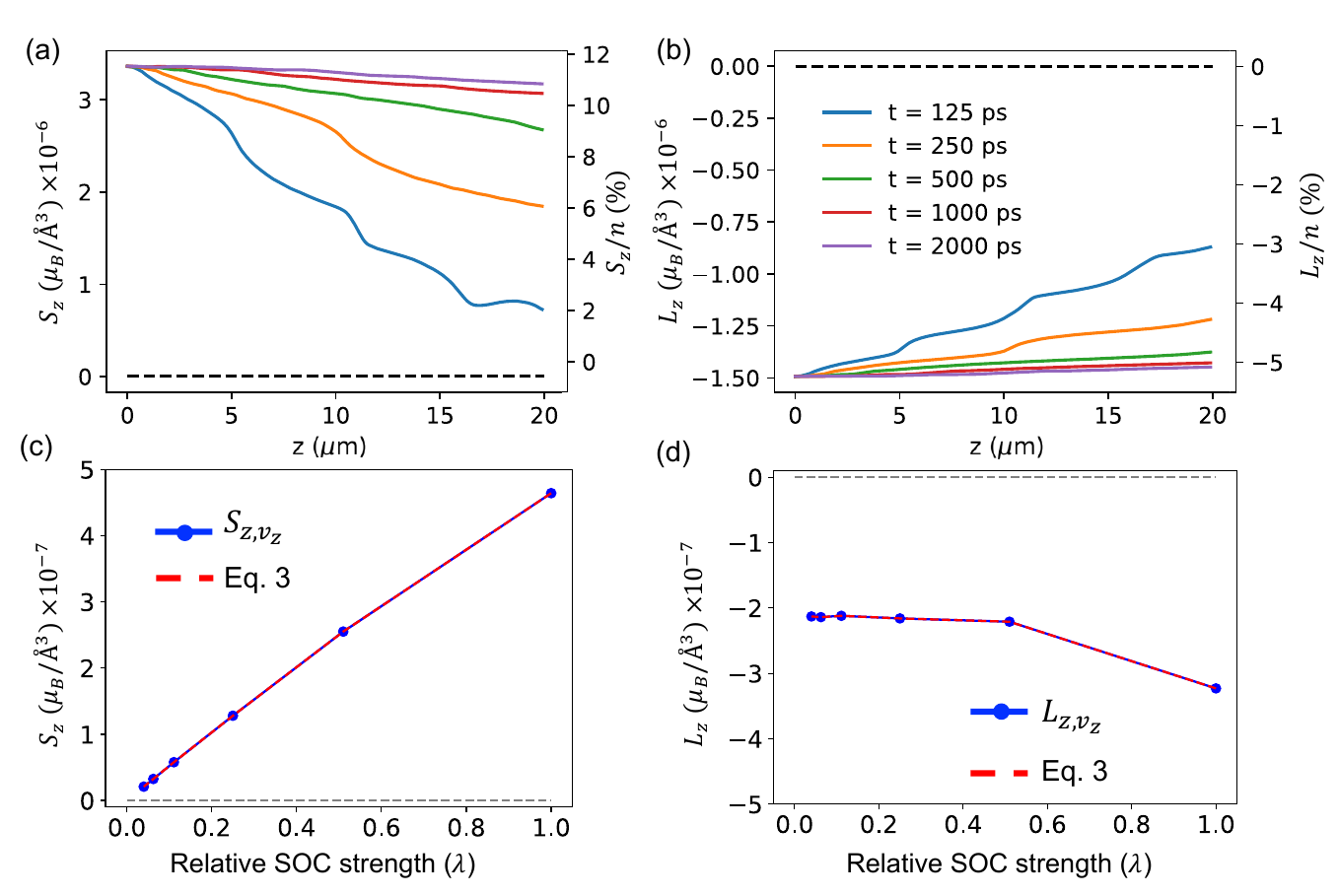}
\caption{Coherent transport dynamics in the right-handed Se enantiomer. Spatial dynamics under coherent charge transport along the chiral axis $\hat{z}$. (a-b) Spatial distribution of longitudinal spin and orbital polarization density components at different time steps. Both spin and orbital polarizations are represented as density $Tr(\hat{\rho} \hat{S})$ and $Tr(\hat{\rho} \hat{L})$, respectively, in atomic units (left scale) and as percentages $\frac{Tr(\hat{\rho}\hat{S})}{Tr(\hat{\rho})} \times$ 100\% and $\frac{Tr(\hat{\rho}\hat{L})}{Tr(\hat{\rho})} \times$ 100\%, respectively, (right scale). The latter definition is effectively the same as  $P_s = (\rho_{n\uparrow} - \rho_{n\downarrow})/(\rho_{n\uparrow} + \rho_{n\downarrow})\times$100\%, but avoiding the difficulties of defining up and down spins with SOC.  
%
%
(c-d) Spin and orbital polarization densities as a function of SOC strength. Here, SOC strength ($\lambda$) is represented relative to the natural value $\lambda_0$ of bulk Se. The real-time FPDMD results without scatterings (solid blue) agree with analytical solutions from Eq.~\ref{Eq:coherent-magnetization} (dashed red).}
\label{fig:co-transport}
\end{figure}

In the coherent (ballistic) regime, we propagate the density matrix via 
Eq.~(\ref{Eq:lindblad}) with $\mathcal{L}=0$. 
Unpolarized carriers are injected 
along the chiral axis $\hat{z}$ from a contact at $z=0$, driven by a 
chemical potential difference $\Delta\mu$ between source and drain. 
Spin and orbital angular momenta are evaluated along $z>0$ at multiple 
time steps; results are shown in Fig.~\ref{fig:co-transport}.

Figs.~\ref{fig:co-transport}(a,b) show the spatiotemporal evolution of 
longitudinal spin ($S_z$) and orbital ($L_z$) polarization, reaching 
steady state after $\sim$2~ns. Both are aligned with the chiral axis 
$\hat{z}$; transverse components are negligible. At $\Delta\mu = 0.1$~eV 
with $\mu$ at the conduction band minimum, the steady-state spin and 
orbital polarization reach $\sim$12\% and $\sim$5\%, respectively. The 
opposite enantiomer yields equal and opposite polarizations by symmetry 
(Fig.~S4,~\cite{SI}).

In the coherent limit (Lindbladian scattering operator {$\mathcal{L}=0$), Eq.~(\ref{Eq:lindblad}) gives the following analytical expression for the magnetization ($M^{b}$) (see SI Section~IV):
%
%
\begin{multline}
M^b = \sum_{\bf k}\sum_{n}
\Bigl[f_{\mu_R,T}(\varepsilon_{n}({\bf k})) \\
-\,\theta(v^z_{n}({\bf k}))
\bigl(f_{\mu_R,T}(\varepsilon_{n}({\bf k}))
- f_{\mu_L,T}(\varepsilon_{n}({\bf k}))\bigr)
\Bigr]S_{n}^b({\bf k}),
\label{Eq:coherent-magnetization}
\end{multline}
where R/L denote the right (drain) and left (source) contacts, 
$\theta(v^z_n({\bf k})) = 1$ for states traveling source to drain 
($v^z_n > 0$) and zero otherwise, and $S_n^b({\bf k})$ is the diagonal 
spin matrix element in the Bloch basis ($b = x,y,z$).
As shown in Figs. 2(c,d), FPDMD agrees exactly with 
Eq.~(\ref{Eq:coherent-magnetization}), confirming that spin 
polarization generated in coherent charge transport is determined entirely by the 
spin polarized occupation difference at the source and drain.

The role of SOC dependence of $S_z$ and $L_z$ is analyzed by varying the strength of the SOC ($\lambda$). $\lambda$ is tuned by scaling the speed of light parameter in the relativistic pseudopotentials employed in first-principles calculations~\cite{HamannONCV}. As seen in Fig.~\ref{fig:co-transport}(c), no spin accumulation is observed in the absence of SOC. This is due to the complete cancellation of the up$-$ and down$-$spin components of two bands at each ${\bf k}-$point. In contrast to spin, both degenerate bands have the same orbital angular momentum regardless of the value of $\lambda$, and thus orbital polarization remains mostly insensitive to SOC (see Fig.~\ref{fig:co-transport}(d)). This observation is consistent with the proposed mechanism \cite{Hiroshi-orb} that low SOC materials generate mostly OAM but converts to spin at the interface with the leads \cite{REE_prl2015,CISSAdhikari}. 
We also analyzed the effect of structural chirality on spin polarization under the coherent transport condition. As shown in section-V of SI, increasing structural chirality $S^2(u)$ leads to a monotonic enhancement of the longitudinal spin $S_{z,v_z}$ and orbital $L_{z,v_z}$ polarization (see Figs. S5 and S6 of the SI and \cite{AndrewCCM}).

\textit{Incoherent dynamics with a relaxation time approximation:}
To quantify the collinear Edelstein effect, we evaluate the standard 
linear-response magnetization~\cite{CEE_Souza} within a 
relaxation-time approximation $\tau_c$, benchmarked against FPDMD with 
the Lindbladian replaced by 
$\left.\frac{\partial\rho}{\partial t}\right|_{\mathrm{scatt}} = 
-(\rho(t)-\rho_0)/\tau_c$. Under an applied field $E_z\parallel\hat{z}$, 
this gives~\cite{CEE_Souza},
%
\begin{equation}
\label{Eq:RTA-magnetization}
M^z = \frac{e\tau_c}{\hbar}\sum_a E_a\,{\bf K}_{az},
\end{equation}
where the magnetoelectric tensor ${\bf K}_{ab}$~\cite{CEE_Souza} is
\begin{equation}
{\bf K}_{ab} = \sum_n\sum_{\bf k}
v_n^a({\bf k})\,m_n^b({\bf k})
\left(-\frac{\partial f_{n{\bf k}}^0}{\partial\varepsilon_n({\bf k})}\right),
\end{equation}
with $v_n^a=\partial_{k_a}\varepsilon_n$, $m_n^b$ the spin or orbital
magnetic moment, and $f^0_{n{\bf k}}$ the equilibrium Fermi-Dirac
distribution (see SI Sec.~VI for the full derivation).
The carrier lifetime $\tau_c\approx 475$~fs at 50~K at $\mu=0.1$~eV (reference to CBM) is
extracted from FPDMD with e-ph
couplings~\cite{Xu2020,Xu2021,Habib2022,Xu2024-lw,Xu2024-ze}.

\begin{figure}[h]
\includegraphics[width=1.0\columnwidth]{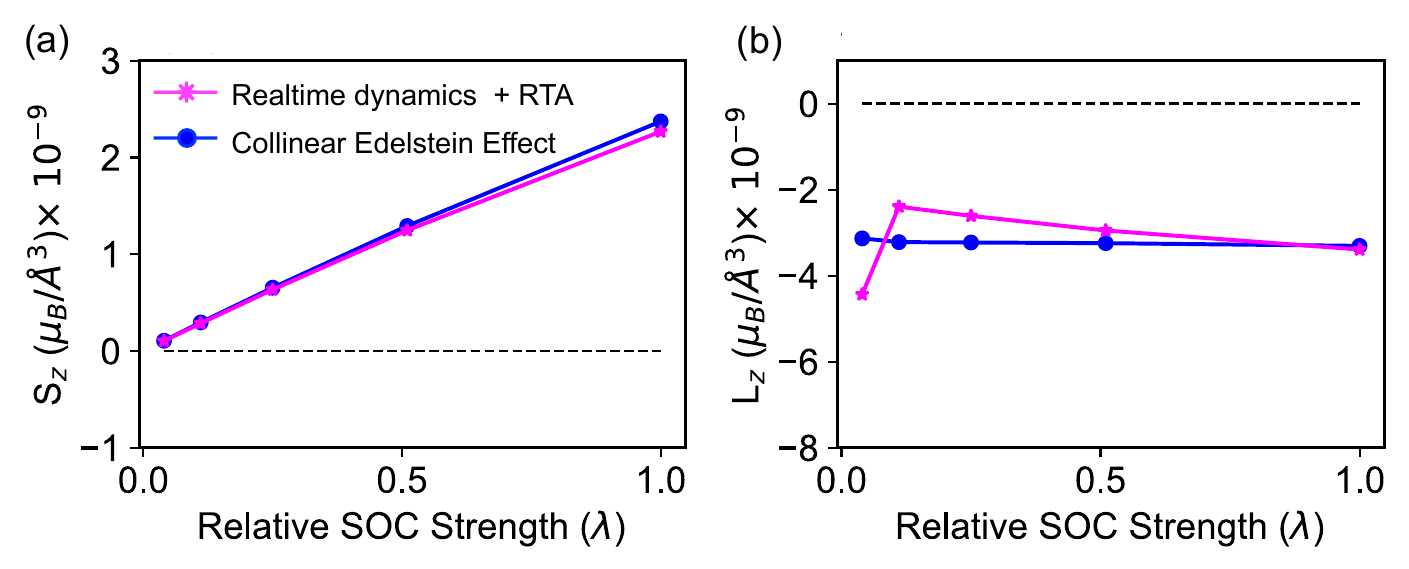}
\caption{Dependence of the induced spin and orbital polarization on the relative SOC strength ($\lambda$) under RTA and CEE. (a) Spin ($S_z$) and (b) orbital polarization ($L_z$) obtained from real-time density-matrix dynamics under RTA, compared with the CEE using Eq.~(\ref{Eq:RTA-magnetization}) derived from semi-classical BTE. The two results agree well across varying spin-orbit coupling strengths. Here, SOC strength ($\lambda$) is given relative to its bulk Se value $\lambda_0$.
}
\label{fig:RTA}
\end{figure}

Fig.~\ref{fig:RTA}(a) shows that both approaches yield a monotonic increase
of spin polarization with SOC strength, consistent with spin--momentum
locking generating a longitudinal spin accumulation under bias, while
orbital polarization shows only a weak SOC dependence
[Fig.~\ref{fig:RTA}(b)].
The weak SOC dependence of orbital polarization suggests it may drive CISS in low-SOC materials, where OAM converts to spin at the interface with high-SOC metallic leads~\cite{REE_prl2015,CISSAdhikari}.
Although RTA captures spin accumulation from spin--momentum locking, it
assumes spin-independent scattering and cannot reproduce the
experimentally observed length-dependent increase of spin
polarization~\cite{science_dna,Valy2019,Mishra2020,Mondal2016}. We
therefore incorporate explicit e--ph interactions with self-consistent SOC
in the Lindbladian operator[Eq.~(\ref{Eq:lindblad})]~\cite{Xu2024-ze,Xu2021,simoni2025}.

\begin{figure}
\includegraphics[width=1.0\columnwidth]{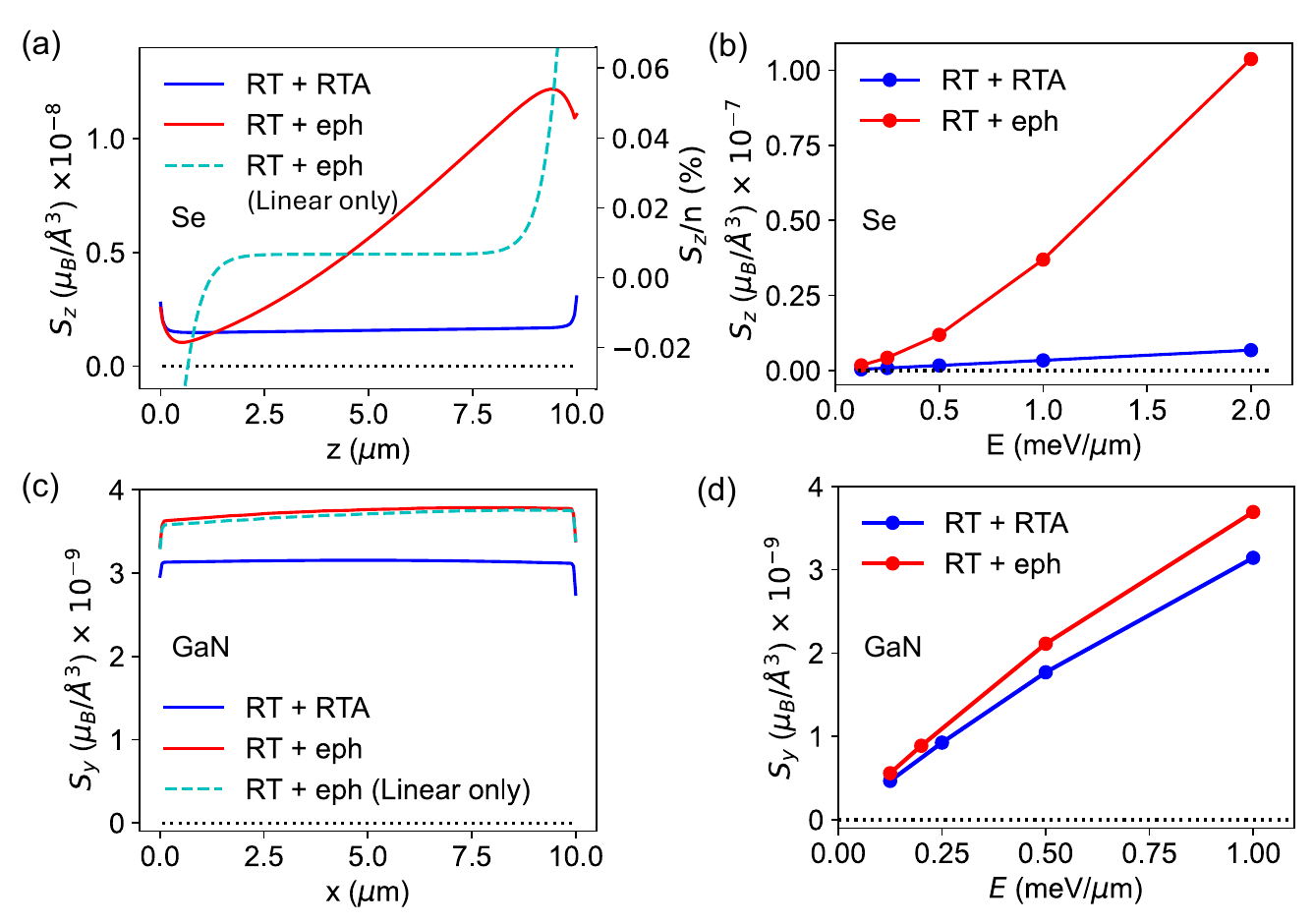}
\caption{Generation of spin polarization for Se and GaN during incoherent
transport with explicit electron-phonon scatterings.
(a) Steady-state spin polarization under RTA (blue), full e-ph scattering
(red), and e-ph with linear contribution only (cyan) for
$\vec{E}=0.5$~meV/$\mu$m.
(b) $S_z$ vs.\ applied field near $z=10~\mu$m: $S_z\propto E^2$ for
explicit e-ph and $S_z\propto E$ for RTA.
(c) Same calculation for GaN.
(d) $S_z$ vs.\ $E$ for GaN; $S_z\propto E$ in both cases, consistent
with the Rashba-Edelstein effect.
}
\label{fig:incoherent}
\end{figure}


\textit{Incoherent dynamics with explicit electron-phonon scatterings:}
Fig.~\ref{fig:incoherent}(a) shows the steady-state spin polarization
($\mu=0.1$~eV above CBM, $\Delta\mu=5$~meV, $T=50$~K). Within
RTA (blue curve), $S_z$ is spatially uniform; explicit e--ph scattering  (red curve) instead
produces a position-dependent accumulation with $S_z$ growing
approximately linearly with length. Computing $S_z$ vs.\ applied field
$E$ [Fig.~\ref{fig:incoherent}(b)] reveals $S_z\propto E$ under RTA
[consistent with Eq.~(\ref{Eq:RTA-magnetization})] but $S_z\propto E^2$
under full e--ph dynamics. Expanding $\rho=\rho_0+\delta\rho$, where $\delta\rho$ is the nonequilibrium correction
induced by the applied field,  and
substituting into the Lindblad operator [Eq.~(\ref{eq:Lindblad})] yields
the second-order contribution
\begin{multline}
\mathcal{L}^{(2)}[\rho_0+\delta\rho]_{n_1n_2}^k =
\frac{2\pi}{\hbar N_q}\sum_{q\lambda,\pm}\sum_{n'k'}
n_{\bm{q}\lambda}^{\pm}\mathrm{Re}\Big[
-\delta f^{k}_{n_1}A^{{\bf q}\lambda\pm}_{{\bf k}n_1;{\bf k}'n'}\\
\delta f^{k'}_{n'}A^{{\bf q}\lambda\mp}_{{\bf k}'n';{\bf k}n_2}
+A^{{\bf q}\lambda\mp}_{{\bf k}'n_1;{\bf k}n'}
\delta f^{k'}_{n'}A^{{\bf q}\lambda\pm}_{{\bf k}n';{\bf k}'n_2}
\delta f^{k}_{n_2}\Big],
\label{eq:L2}
\end{multline}
which scales as $(\delta f)^2\propto E^2$. Retaining only the linear term
recovers the spatially uniform, $S_z\propto E$ profile of RTA
[Fig.~\ref{fig:incoherent}(a), cyan], confirming that the nonlinear buildup arises from second-order contribution from spin-dependent e-ph scattering (Eq.~\ref{eq:L2}). 
For comparison, we perform the same calculations for achiral wurtzite 
GaN~\cite{SI}, which has broken inversion but mirror symmetry, and 
therefore no structural chirality. GaN exhibits a linear spin response 
($S_{\textcolor{red}{y}} \propto E$) under both RTA and explicit e-ph scattering 
[Figs.~\ref{fig:incoherent}(c,d)], consistent with the Rashba-Edelstein 
effect. The absence of nonlinear spin accumulation in GaN confirms that 
the $S_z \propto E^2$ response in Se is a direct consequence of 
structural chirality, not simply of broken inversion symmetry.
%

\begin{figure}[h]
\includegraphics[width=1.0\columnwidth]{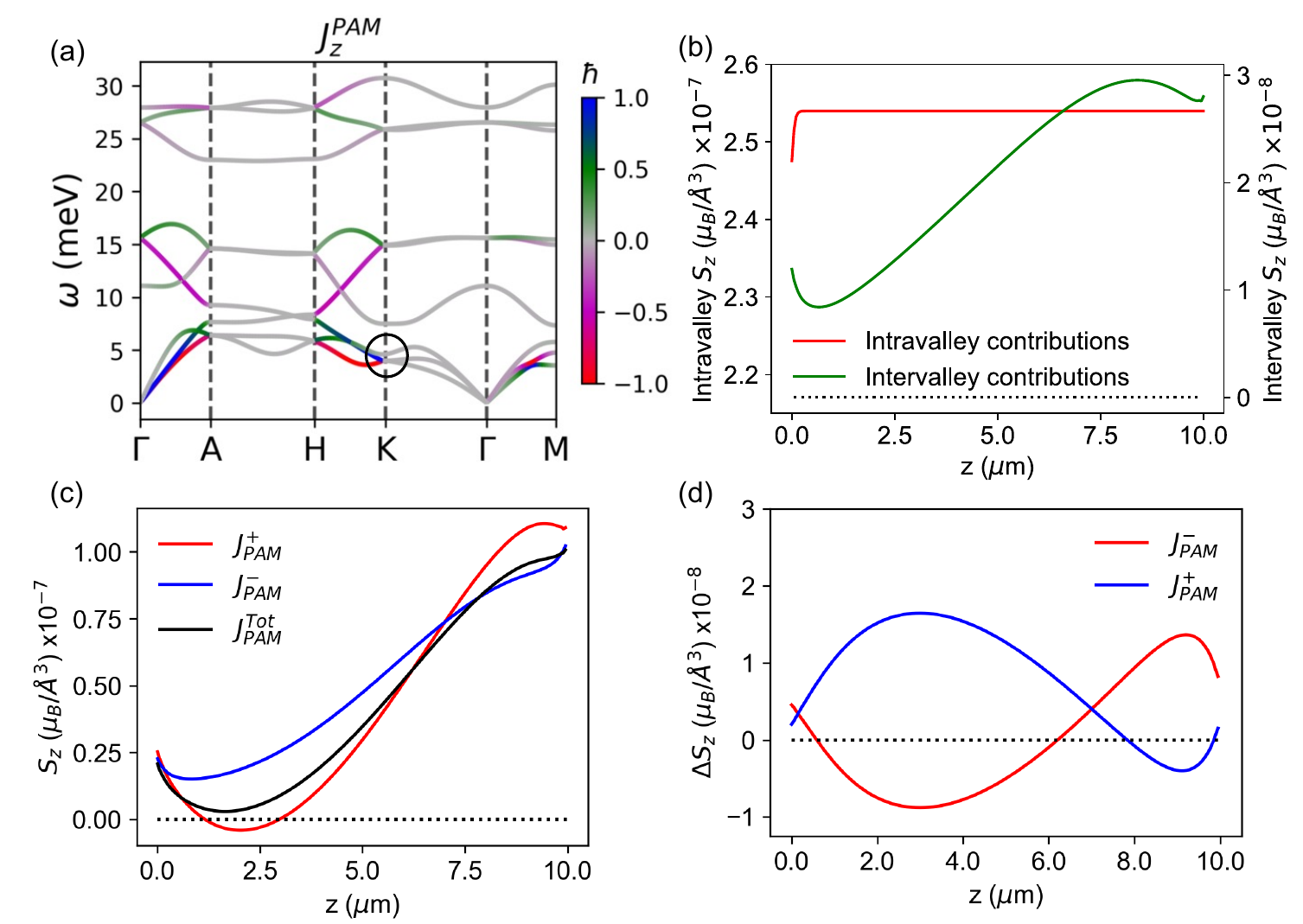}
\caption{Contribution of e–ph scattering into
intravalley and intervalley channel. (a) Phonon band structure of Se projected onto the
$z$-component of phonon angular momentum ($J_z^{PAM}$).
(b) Intervalley and intravalley e-ph contributions to spin polarization
(black circles).
(c) Spin polarization from positive ($J_{PAM}^+$) and negative
($J_{PAM}^-$) PAM channels of the first acoustic phonon mode; for $J_{PAM}^{Tot}$ the Linbladian scattering is scaled by 0.5 for
comparison with the partitioned phonon channels ($J_{PAM}^+$ and $J_{PAM}^-$) since they only cover half the phonon phase space.
  (d) The change in $S_z$ for the different phonon channels compared to $J_{PAM}^{Tot}$ in panel (c) where $\Delta S_z(J_{PAM}^{\pm}) = S_z(J_{PAM}^{\pm}) - S_z(J_{PAM}^{Tot})$.
}
\label{fig:PAM}
\end{figure}

To identify the microscopic origin of the spatial spin buildup in chiral Se, we decompose e--ph scattering into intravalley and intervalley channels [Fig.~\ref{fig:PAM}(b)]. The intravalley contribution is spatially uniform, while the intervalley contribution grows with distance and dominates, establishing intervalley scattering as the primary driver. 
Phonon angular momentum (PAM) plays an additional role in the 
intervalley scattering: chiral phonons along $\Gamma\to A$ and 
$H\to K$ [Fig.~\ref{fig:PAM}(a)], particularly the low-energy 
acoustic modes near $K$ (circled), mediate this process at 50~K. 
We partition the first-acoustic-mode e--ph scattering into positive 
and negative PAM channels, $J_{PAM}^+$ and $J_{PAM}^-$; 
$J_{PAM}^{Tot}$ is obtained with the Lindbladian scaled by 0.5 for 
a fair per-channel comparison.
%
Fig.~\ref{fig:PAM}(c,d) shows the total contribution exhibits the characteristic spatial increase of spin polarization; $J_{\mathrm{PAM}}^{+}$ slightly increases the slope of spin accumulation, while $J_{\mathrm{PAM}}^{-}$ has the opposite effect. 
Fig.~\ref{fig:PAM}(d) shows $\Delta S_z(J_{PAM}^{\pm}) = S_z(J_{PAM}^{\pm}) - S_z(J_{PAM}^{Tot})$
asymmetry between opposite PAM channels. The average of $\Delta S_z$ for $J_{PAM}^{\pm}$ is nonzero and reflects 
a net angular momentum transfer from chiral phonons to electrons:
nonequilibrium transport breaks time-reversal
symmetry, producing unequal valley occupations and asymmetric scattering
for phonons with opposite chirality.

A sharp drop in $S_z$ near both boundaries reflects spin relaxation from
SOC and e-ph scattering upon charge injection, consistent with
experiment~\cite{Valy2019}, where a similar
pattern of increasing spin accumulation with longer
lengths followed by an initial drop close to the boundaries
has been observed in 2D chiral perovskites. To confirm spin relaxation near the boundaries, we performed FPDMD transport simulations with spin injection, where a spin diffusion length $L_s$ of $\sim$ 0.24 $\mu$m is extracted, qualitatively in agreement with the length scale of initial spin polarization drop in charge injection shown in Fig.~\ref{fig:incoherent}(a) (details see Fig.~S7 in SI Sec.~IX).  
E--ph scattering thus simultaneously relaxes spin on short length scales
and generates it on longer ones; their competition sets the net CISS
signal.

To connect with experiment, we compare against Te single crystals, 
which are isostructural with Se but lack direct spin transport 
measurements in Se. Shalygin \textit{et al.}~\cite{Shalygin2012} 
report $\sim$1.4\% current-induced spin polarization for holes in Te 
at 77~K and current density $J = 1{,}400$~A/cm$^2$ with a carrier concentration of 
$4\times10^{16}$~cm$^{-3}$. Under matched conditions, 
our Se calculations for electron carriers yield $\sim$2\%, in qualitative 
agreement given the differences in carrier type and SOC strength
(see Fig.~S11 and SI Section~XII).
%
%
%
%
Another important experimental comparison is length-dependent spin accumulation seen in ~\cite{Valy2019,Mishra2020} which is captured by our incoherent 
transport calculations and is attributed to asymmetric, 
spin-dependent intervalley scattering mediated by chiral phonons. Importantly, the spatial increase in spin polarization is only captured here because both spatial resolution and explicit e--ph scattering are included, going beyond previous simulations.

In summary we present a unified first-principles framework---combining density-matrix dynamics, a Wigner-function formalism, and Lindblad electron-phonon scattering---to resolve the microscopic origin of CISS in a prototypical chiral solid - trigonal selenium. A central finding is the quantitative distinction between CISS and the collinear Edelstein effect: CEE produces a spatially uniform spin polarization scaling linearly with applied field (S$_z$ $\propto$ E), while CISS generates a length-dependent spin accumulation with a nonlinear field dependence (S$_z$ $\propto$ E$^2$), arising from second-order spin-dependent intervalley  scattering mediated by chiral phonons. Additionally, unlike spin, orbital 
polarization is largely insensitive to SOC strength. These results establish electron–phonon interactions-not spin–orbit coupling alone-as the essential ingredient for CISS, and provide quantitative guidance for engineering spin selectivity in nonmagnetic chiral materials.

\textbf{Acknowledgments} We thank Aron W. Cummings and Michael E. Flatt\'e for helpful discussions. 
The authors acknowledge support for the theoretical and code development of first-principles spatiotemporal density-matrix dynamics by the computational chemical science program within the Office of Science at DOE under grant No. DE-SC0023301. The authors acknowledge the support for materials theory application as part of the Center for Hybrid Organic-Inorganic Semiconductors for Energy (CHOISE), an Energy Frontier Research Center funded by the Office of Basic Energy Sciences, Office of Science within the US Department of Energy (DOE).
Calculations were carried out at the National Energy Research Scientific Computing Center (NERSC), a U.S. Department of Energy Office of Science User Facility operated under Contract No. DEAC02-05CH11231.

\bibliography{main}

\end{document}